\documentclass[a4paper,american,british,a4paper,fleqn,usenatbib]{mnras}
\usepackage[T1]{fontenc}
\usepackage[latin9]{inputenc}
\usepackage{url}
\usepackage{amsmath}
\usepackage{amssymb}
\usepackage{graphicx}
\usepackage[authoryear]{natbib}

\makeatletter

\providecommand{\tabularnewline}{\\}

\usepackage{mathptmx}
\usepackage{txfonts}

\usepackage[T1]{fontenc}
\usepackage{ae,aecompl}

\title[New regime of steady-state nuclear burning]{Carbon production on accreting neutron stars in a new regime of stable nuclear burning}
\author[L. Keek \& A. Heger]{L. Keek$^1$\thanks{E-mail: l.keek@gatech.edu} and A. Heger$^{2,3,4}$ \\
$^1$Center for Relativistic Astrophysics, School of Physics, Georgia Institute of Technology, 837 State Street, Atlanta, GA 30332-0430, USA\\
$^2$Monash Center for Astrophysics, School of Physics and Astronomy, Monash University, Victoria, 3800, Australia\\
$^3$Shanghai Jiao-Tong University, Center for Nuclear Astrophysics, Department of Physics and Astronomy, Shanghai 200240, P.~R.~China\\
$^4$University of Minnesota, School of Physics and Astronomy, Minneapolis, MN 55455, USA
}

\date{Accepted XXX. Received YYY; in original form ZZZ}
\pubyear{2015}

\makeatother

\usepackage{babel}
\begin{document}
\label{firstpage}
\pagerange{\pageref{firstpage}--\pageref{lastpage}}
\maketitle
\begin{abstract}
Accreting neutron stars exhibit Type I X-ray bursts from both frequent
hydrogen/helium flashes as well as rare carbon flashes. The latter
(superbursts) ignite in the ashes of the former. Hydrogen/helium bursts,
however, are thought to produce insufficient carbon to power superbursts.
Stable burning could create the required carbon, but this was predicted
to only occur at much larger accretion rates than where superbursts
are observed. We present models of a new steady-state regime of stable
hydrogen and helium burning that produces pure carbon ashes. Hot CNO
burning of hydrogen heats the neutron star envelope and causes helium
to burn before the conditions of a helium flash are reached. This
takes place when the mass accretion rate is around $10\,\%$ of the
Eddington limit: close to the rate where most superbursts occur. We
find that increased heating at the base of the envelope sustains steady-state
burning by steepening the temperature profile, which increases the
amount of helium that burns before a runaway can ensue.\end{abstract}
\begin{keywords}
accretion, accretion disks --- methods: numerical --- nuclear reactions,
nucleosynthesis, abundances --- stars: neutron --- X-rays: binaries
--- X-rays: bursts
\end{keywords}

\section{Introduction}

 In low-mass X-ray binaries hosting a neutron star, hydrogen- and
helium-rich material may be transferred from the companion star to
the neutron star by Roche-lobe overflow. Compression due to a high
surface gravity induces nuclear fusion, giving rise to a broad range
of observed phenomena. Most prominent are Type I X-ray bursts with
typical durations of $10-100\,\mathrm{s}$ produced by runaway thermonuclear
burning \citep[see also \citealt{Lewin1993,Strohmayer2006}]{Grindlay1976,Woosley1976,Maraschi1977}.
 Stable burning is inferred from the absence of bursts at mass accretion
rates above $\sim10\,\%-30\,\%$ of the Eddington limit \citep{Paradijs1988,Cornelisse2003}.
In that case hydrogen and helium burn continuously in a steady state,
without triggering a thermonuclear runaway. All these burning regimes
are reproduced by theory \citep{Fujimoto1981}, but there is a large
mismatch in the conditions required for certain regimes to occur.
For example, bursts are predicted to persist up to $\sim100\,\%$
of the Eddington limit \citep[e.g., ][]{Heger2005,Keek2009,Zamfir2014}.

 Many subsequent X-ray bursts or long periods of stable burning of
hydrogen and helium burning leave behind an ashes layer that is rich
in carbon. Runaway carbon fusion in these ashes is thought to power
the rare ``superbursts'': hours-long flashes from the same X-ray
sources \citep{Cornelisse2000,Cumming2001,Strohmayer2002}. All superbursting
neutron stars also exhibit short bursts, but superburst observations
constrain the carbon mass fraction to be larger \citep[$\sim 20\,\%$; ][]{Cumming2006}
than what hydrogen/helium bursts are thought to produce \citep[$\lesssim 5\,\%$; e.g.,][]{Woosley2004}.
Chemical separation could increase the carbon fraction \citep{Medin2011},
but would increase the recurrence times. These sources have relatively
low burst rates \citep{Zand2004}, suggesting that aside from bursts
some stable burning occurs. This may produce larger amounts of carbon
\citep[e.g.,][]{Stevens2014}, but no such stable burning mode was
predicted by theory.

In this Letter we investigate the ignition conditions of hydrogen
and helium burning, and we identify a new stable burning regime at
mass accretion rates bordering the burst regimes. Furthermore, time-dependent
multi-zone models show that this regime is more prevalent when the
base heating of the neutron star envelope is strong.

\section{Numerical methods}

\subsection{One-zone model}

\label{sub:One-zone-Model}

We construct a simple one-zone numerical model to study the ignition
conditions of hydrogen and helium burning as well as the stability
of the burning processes in the neutron star envelope. The zone is
given a temperature, $T$, and density, $\rho$, as well as mass fractions
$X$, $Y$, $Z$, for $^{1}\mathrm{H}$, $^{4}\mathrm{He}$, and the
combined CNO isotopes, respectively. We use a solar accretion composition
of $X=0.73$, $Y=0.25$, and $Z=0.02$, and we assume all accreted
CNO is initially $^{14}\mathrm{N}$. The Helmholtz equation of state
and opacity routines \citep{2000TimmesEOS,Timmes2000kap} are employed
to calculate the pressure, $P$, and opacity, $\kappa$. The column
depth is derived as $y=P/g$. We use a local gravitational acceleration
$g=1.86\times10^{14}\,\mathrm{cm\,s^{-2}}$, corresponding to a neutron
star with a gravitational mass of $1.4\,M_{\odot}$ and a radius of
$10\,\mathrm{km}$ in the Newtonian case or $11.2\,\mathrm{km}$ for
General Relativity \citep[e.g.,][]{Keek2011}.

Radiative cooling is implemented with a specific rate of \citep[e.g.,][]{Fujimoto1981}:
\begin{equation}
\epsilon_{\mathrm{cool}}=\frac{acT^{4}}{3\kappa y^{2}}.\label{eq:ecool}
\end{equation}
We calculate the specific energy generation rate from nuclear burning,
$\epsilon_{\mathrm{nuc}}$. For helium burning we use the $3\alpha$
rate from \citet{Fynbo2005}. For hydrogen burning we employ the hot
$\beta$-limited CNO cycle ($\beta$CNO) for temperatures $T>0.8\times10^{8}\,\mathrm{K}$
following \citet{Bildsten1998}, and at lower temperatures the CNO
cycle using the $\mathrm{^{14}N(p,\gamma)}\mathrm{^{15}O}$ rate as
the slowest part of the cycle following \citet{Kippenhahn1994}. Screening
is implemented following \citet{GraboskeScreening}. Finally, the
base of the envelope is heated by pycnonuclear and electron-capture
reactions in the neutron star crust \citep{Haensel1990,Haensel2003,Gupta2007,Schatz2014Nature},
or possibly by the dissipation of rotational energy at the bottom
of the ocean \citep{Inogamov2010}. The amount of heat generated per
accreted nucleon, $Q_{\mathrm{b}}$, is treated as a free parameter,
and we include base heating as a specific rate: $\epsilon_{\mathrm{b}}=Q_{\mathrm{b}}\dot{m}/y$,
where $\dot{m}$ is the specific mass accretion rate. We use a typical
value for $Q_{\mathrm{b}}$ of $0.1\,\mathrm{MeV\,u^{-1}}$ \citep[e.g.,][]{Cumming2006}.
The total heating rate is $\epsilon_{\mathrm{heat}}=\epsilon_{\mathrm{nuc}}+\epsilon_{\mathrm{b}}$.
We assume the (small) contribution from compressional heating to be
part of $Q_{\mathrm{b}}$ \citep[e.g.,][]{Fujimoto1981}.

In the steady-state case, we determine $\dot{m}$ by equating the
timescales for burning and accretion, $\tau_{\mathrm{nuc}}=\tau_{\mathrm{acc}}$,
with $\tau_{\mathrm{acc}}\equiv y/\dot{m}$: 
\begin{equation}
\dot{m}=y/\tau_{\mathrm{nuc}}.\label{eq:mdot}
\end{equation}
 For helium burning $\tau_{\mathrm{nuc}}\equiv E_{\mathrm{nuc}}Y/\epsilon_{\mathrm{nuc}}$,
where $E_{\mathrm{nuc}}=6.97\times10^{18}\,\mathrm{erg\,g^{-1}}$
is the $3\alpha$ energy yield per unit mass; for hydrogen burning
through the $\beta$CNO cycle, $\tau_{\mathrm{nuc}}$ is set by the
sum of the half lives of $^{14}\mathrm{O}$ and $^{15}\mathrm{O}$,
$\tau_{\beta}=192.86\,\mathrm{s}$: $\tau_{\mathrm{nuc}}\equiv\tfrac{14X}{4Z}\tau_{\beta}$,
where $14$ accounts for the mass number of $^{14}\mathrm{N}$, and
$4$ is the net number of protons that are burned in each cycle. We
express $\dot{m}$ as a fraction of the Eddington limited value for
solar composition: $\dot{m}_{\mathrm{Edd}}=8.78\times10^{4}\,\mathrm{g\,s^{-1}}$.

\subsection{Multi-zone time-dependent models}

\label{sub:Multi-Zone-Time-Dependent-Models}

We employ the one-dimensional multi-zone hydrodynamics code KEPLER
\citep{Weaver1978} to create time-dependent models of the neutron
star envelope. The version of KEPLER and the setup of the models is
similar to a previous study by \citet{Keek2014} \citep[see also][]{Woosley2004,Heger2005,Heger2007}.
We refer to these studies for details of the code, and here we summarize
the main characteristics of our models. The code includes a large
nuclear network using thermonuclear reaction rates from the REACLIB
$2.0$ compilation \citep{Cyburt2010}. The network includes all CNO
and $\beta$CNO cycle reactions, the $3\alpha$ process using the
rate from \citet{Caughlan1988},\footnote{Within the temperature range considered in our study, the \citet{Caughlan1988}
and \citet{Fynbo2005} $3\alpha$ rates differ at most $4\,\%$.} as well as many other reactions such as break-out from the $\beta$CNO
cycle. We resolve the neutron star envelope in the radial direction.
Material of solar composition is accreted onto a $2\times10^{25}\,\mathrm{g}$
iron substrate, which acts as a thermal buffer into which heat from
nuclear burning can dissipate. Crustal heating is implemented as an
inflowing heat flux at the inner zone. The same accretion composition
and neutron star parameters are used as for the one-zone model (Section~\ref{sub:One-zone-Model}).
The results in this Letter are presented without GR corrections or
redshifts \citep[see][]{Keek2011}.

\section{Results}

\subsection{Ignition conditions}

\begin{figure}
\includegraphics{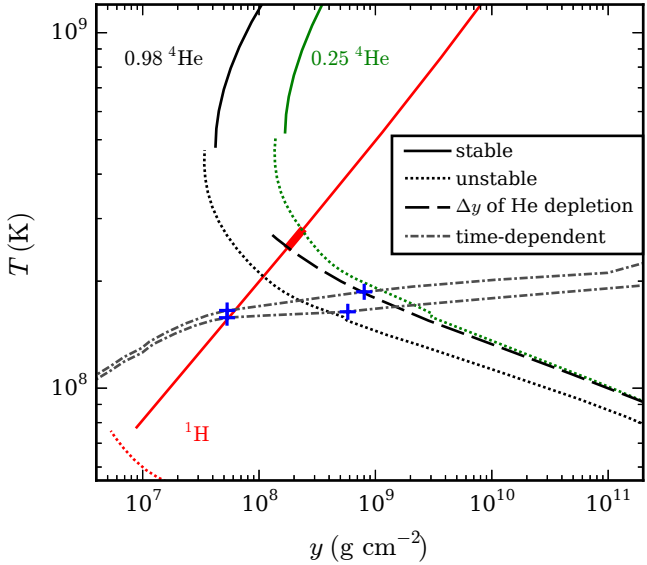}

\caption{\selectlanguage{american}%
\label{fig:stability}\foreignlanguage{british}{Column depth, $y$,
as a function of temperature, $T$, for stable and unstable burning
of H and He from a solar accretion composition, as well as a He mass
fraction of $0.98$ when all H has burned to He. Accreted material
moves to larger $y$ and $T$ with time, and thus reaches any of the
ignition curves. Starting at the stable H line, stable burning depletes
He after $\Delta y$: the He depletion depth is the stable H line
\emph{plus} $\Delta y$. The thick solid line indicates where the
new stable burning regime of H and He occurs when $\Delta y<$stable
H line. The temperature profiles for two time-dependent models are
shown: steady-state burning (upper line; $Q_{\mathrm{b}}=0.75\,\mathrm{MeV\,u^{-1}}$)
and just prior to unstable He ignition (lower; $Q_{\mathrm{b}}=0.1\,\mathrm{MeV\,u^{-1}}$).
Their respective locations of H and He depletion are marked with `+'.}\selectlanguage{british}%
}
\end{figure}
We use the one-zone model to reproduce the ignition conditions of
hydrogen and helium burning derived by \citet{Fujimoto1981}. Steady-state
burning requires a balance between heating and cooling: $\epsilon_{\mathrm{heat}}=\epsilon_{\mathrm{cool}}$.
For a range of temperatures, we determine the column depth where this
condition is met for hydrogen and helium burning separately (``stable''
in Fig.~\ref{fig:stability}), and during stable burning the respective
element is depleted at that location. A thermonuclear runaway ensues
when the heating rate increases faster with $T$ than the cooling
rate. Both for hydrogen and helium burning we locate the boundary
of the unstable region (``unstable'' in Fig.~\ref{fig:stability}):
\begin{equation}
\frac{d\epsilon_{\mathrm{heat}}}{dT}=\frac{d\epsilon_{\mathrm{cool}}}{dT}.\label{eq:dnucdcool}
\end{equation}

Material accreted onto a neutron star is compressed and heated over
time, such that it moves to higher $y$ and $T$ in Fig.~\ref{fig:stability}.
The burning behaviour is determined by where the ``stable'' and
``unstable'' lines are reached \citep{Fujimoto1981,Fushiki1987ApJ,Bildsten1998}.
Although CNO burning of hydrogen is unstable at low temperatures,
over most of the considered temperature range hydrogen burning is
stable, because the $\beta$CNO rate is independent of temperature.
The stability of helium burning changes at $T\simeq5\times10^{8}\,\mathrm{K}$
(we take the location to be where $dT/dy=0$; e.g., \citealt{Zamfir2014}).
Furthermore, as the stable hydrogen line crosses the unstable helium
line, helium flashes may occur either before hydrogen is depleted
or after all hydrogen has been burned to helium. Two helium ignition
curves are, therefore, presented: one for the accreted helium fraction
$Y=0.25$ and one where all hydrogen has burned to produce $Y=0.98$.
The transition between pure-helium bursts and mixed hydrogen/helium
bursts is at $T\simeq2\times10^{8}\,\mathrm{K}$.

\subsection{Helium depletion before runaway}

These ignition conditions are calculated assuming that hydrogen and
helium burning are independent. If only helium is accreted, the temperature
increases with depth until ignition is reached. Because of the strong
temperature dependence of the $3\alpha$ rate, very little helium
burns before the ignition depth. If, however, hydrogen burning takes
place at shallow depths, the envelope is heated substantially, and
helium burning is initiated already before its ignition curve is reached.
Therefore, before helium burning dominates the energy generation rate,
some helium already burns away. 

For each point on the stable hydrogen burning line (Fig.~\ref{fig:stability}),
we calculate the $3\alpha$ burning rate assuming all hydrogen has
been converted to helium, and we determine the helium burning timescale,
$\tau_{\mathrm{nuc}}$ (Section~\ref{sub:One-zone-Model}). Given
the $\dot{m}$ associated with steady-state hydrogen burning (Eq.~\ref{eq:mdot}),
we estimate the change in column during $\tau_{\mathrm{nuc}}$ as
$\Delta y=\tau_{\mathrm{nuc}}\dot{m}$. Here we assume that the temperature
remains constant going to larger values of $y$ \citep[e.g.,][]{Fujimoto1981},
and we neglect the small increase in $\epsilon_{\mathrm{nuc}}$ from
the increasing density. Where $\Delta y$ is smaller than the depth
of stable $^{1}\mathrm{H}$ burning (Fig.~\ref{fig:stability}),
helium is depleted at a depth close to where hydrogen burns: both
hydrogen and helium burning proceed in a steady-state, and no thermonuclear
runaway occurs. This is a new regime of stable nuclear burning. It
happens only in a small temperature interval around $T=2.6\times10^{8}\,\mathrm{K}$,
which is associated with a small range of accretion rates $0.08<\dot{m}/\dot{m}_{\mathrm{Edd}}<0.11$
(thick solid line in Fig.~\ref{fig:stability}). At lower $T$ pure
helium bursts ignite, whereas at higher $T$ mixed hydrogen/helium
bursts occur.

This is a simplistic approximation of time-dependent burning. Although
it gives a good qualitative description of the behaviour, time-dependent
models are required for accurate quantitative predictions as well
as for a better understanding of the new stable burning regime.

\subsection{Two time-dependent simulations}

For $\dot{m}=0.1\,\dot{m}_{\mathrm{Edd}}$ and $Q_{\mathrm{b}}=0.1\,\mathrm{MeV\,u^{-1}}$,
where the one-zone model predicts the new stable regime, a KEPLER
model exhibits mixed H/He bursts, indicating the new regime must occur
at somewhat lower $\dot{m}$ for multi-zone models. Therefore, we
create two KEPLER simulations with $\dot{m}=0.02\,\dot{m}_{\mathrm{Edd}}$:
one with the same $Q_{\mathrm{b}}=0.1\,\mathrm{MeV\,u^{-1}}$ as the
one-zone models, and one with stronger base heating of $Q_{\mathrm{b}}=0.75\,\mathrm{MeV\,u^{-1}}$.
As expected from Fig.~\ref{fig:stability}, the first model exhibits
pure helium bursts after hydrogen is depleted \citep[similar to model Zm by][]{Woosley2004}.
The second model, however, displays the new steady-state regime, even
though $\dot{m}$ is much lower than where we expected it to occur.
We compare the ignition conditions --- which exhibit for $Q_{\mathrm{b}}=0.75\,\mathrm{MeV\,u^{-1}}$
only small shifts with respect to $Q_{\mathrm{b}}=0.1\,\mathrm{MeV\,u^{-1}}$
--- to the $T$-profiles of the two models, where for the unstable
model we select a time just prior to the runaway (Fig.~\ref{fig:stability}).
The locations where, respectively, the hydrogen and helium mass fractions
are reduced by an order of magnitude agree well with the one-zone
estimates.

\begin{figure}
\includegraphics{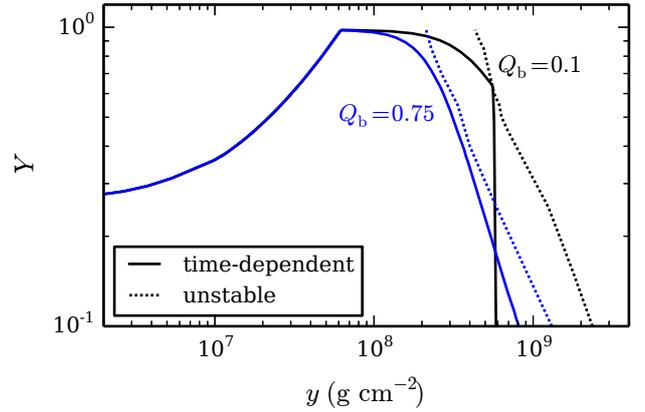}

\caption{\label{fig:race}Helium mass fraction, $Y$, as a function of column
depth, $y$, for two time-dependent models with base heating, $Q_{\mathrm{b}}$,
in units of $\mathrm{MeV\,u^{-1}}$. They overlap at $y<6\times10^{7}\,\mathrm{g\,cm^{-2}}$,
but at larger depth $Y$ drops faster for the hotter model. Dotted
lines mark the conditions for unstable ignition for the two models.
The colder model hits the unstable curve and ignites a burst, whereas
the hotter model does not reach its unstable curve and has stable
burning.}
\end{figure}

Even though the two models have roughly similar temperature profiles,
their burning behaviour is rather different. Up to the depth of hydrogen-depletion,
the helium mass fraction is identical for both models (Fig.~\ref{fig:race}).
For the steady-state model, the temperature profile is slightly steeper
(Fig.~\ref{fig:stability}), and the strong temperature dependence
of the $3\alpha$ reaction causes helium to burn faster than in the
other model. As $Y$ drops, the relevant unstable ignition curve moves
to larger $y$. We create a series of ignition curves for a range
of $Y$, and show their points of intersection with the $T$-profiles
of the KEPLER models (Fig.~\ref{fig:race}). For the unstable model,
 $Y$ is reduced by stable burning to $0.63$ before it hits the
instability line and ignites a burst (Fig.~\ref{fig:race}; see also
\citealt{Woosley2004}). For the steady-state model, however, stable
helium burning proceeds fast enough to avoid its instability line,
and a thermonuclear runaway is never reached.

\begin{figure}
\includegraphics{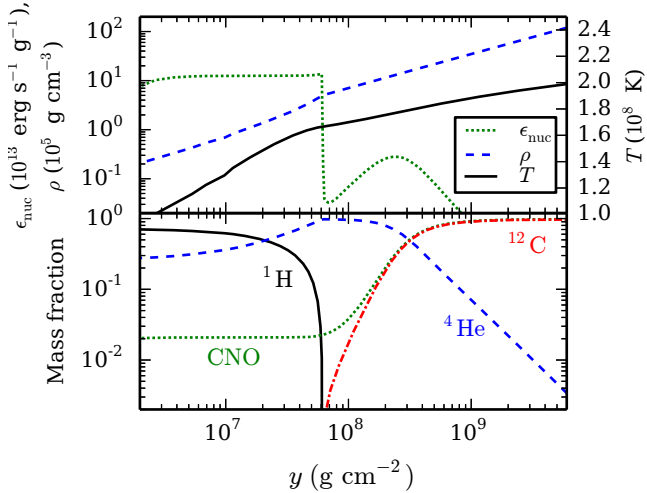}

\caption{\label{fig:profiles}Structure of the neutron star envelope as a function
of column depth, $y$, from a time-dependent multi-zone simulation
of steady-state burning. \textbf{Top}: specific energy generation
rate, $\epsilon_{\mathrm{nuc}}$, density, $\rho$, and temperature,
$T$. \textbf{Bottom}: mass fractions for several isotopes. CNO is
the sum for all C, N, and O isotopes including $^{12}\mathrm{C}$.
At $y\lesssim4\times10^{7}\,\mathrm{g\,cm^{-2}}$ CNO consists mainly
of $^{14}\mathrm{O}$ and $^{15}\mathrm{O}$. }
\end{figure}

We study the steady-state model in more detail. $1.9\,\mathrm{years}$
of accretion and burning were simulated, during which time the burning
layer was replenished $124$ times. The helium mass fraction peaks
at the depth where hydrogen is depleted, and rapidly decreases with
increasing depth before the conditions for runaway $3\alpha$ burning
are reached (Fig.~\ref{fig:profiles}). The temperature in the helium
burning region is below $T=2\times10^{8}\,\mathrm{K}$, which means
that the $^{12}\mathrm{C}(\alpha,\gamma)^{16}\mathrm{O}$ reaction
is very slow compared to $3\alpha$, and $\alpha$-capture is inefficient.
Moreover, because almost all $^{12}\mathrm{C}$ is produced at depths
where hydrogen is depleted, $^{12}\mathrm{C}(p,\gamma)^{13}\mathrm{N}$
does not play a role either. The burning ashes, therefore, consist
exclusively of $98\%$ $^{12}\mathrm{C}$ in addition to the $2\,\%$
accreted CNO mass fraction. The latter are predominantly $^{14}\mathrm{N}$
and $^{15}\mathrm{N}$ from the $\beta$CNO cycle with a ratio of
about $1:2$.

\section{Discussion}

\begin{table}
\begin{centering}
\caption{\label{tab:Theoretical-Nuclear-Burning}Theoretical Nuclear Burning
Regimes$\,^{\mathrm{a}}$}
\begin{tabular}{cl}
\hline 
$\dot{m}/\dot{m}_{\mathrm{Edd}}$ & Burning Regime\tabularnewline
\hline 
 & Deep H flash (burns He)\tabularnewline
$\sim0.1\,\%\,^{\mathrm{b}}$ & \tabularnewline
 & Shallow H flashes and deep He flash \tabularnewline
$0.\,4\%$  & \tabularnewline
 & He flash (stable H burning) \tabularnewline
$8\,\%$ & \tabularnewline
 & \textbf{Stable H/He burning}\tabularnewline
$11\,\%$ & \tabularnewline
 & Mixed H/He flash\tabularnewline
 & \tabularnewline
$\sim100\,\%$$\,^{\mathrm{c}}$ & Marginally stable burning of H/He\tabularnewline
 & \tabularnewline
 & Stable H/He burning\tabularnewline
\hline 
\end{tabular}
\par\end{centering}

$\,^{\mathrm{a}}$ For solar accretion composition and $Q_{\mathrm{b}}=0.1\,\mathrm{MeV\,u^{-1}}$.

$\,^{\mathrm{b}}$ \citet{Peng2007}, including sedimentation.

$\,^{\mathrm{c}}$ \citet{Heger2005}. See also \citet{Keek2009,Zamfir2014,Keek2014}.
\end{table}
We have identified a new steady-state nuclear burning regime on accreting
neutron stars. A range of burning behaviour takes place as a function
of $\dot{m}$, and the new regime occurs in a small $\dot{m}$ interval
close to $0.1\,\dot{m}_{\mathrm{Edd}}$ between the pure helium bursts
and the mixed hydrogen/helium bursts. We summarize the different burning
regimes in Table~\ref{tab:Theoretical-Nuclear-Burning}, although
it should be understood that the boundaries between regimes can be
modified by, e.g., the metallicity \citep{Woosley2004}, the details
of CNO break-out \citep{Fisker2007,Keek2014}, rotational mixing \citep{Piro2007,Keek2009},
and base heating \citep{Keek2009,Zamfir2014}. The small size of the
$\dot{m}$ interval may explain why it has not been identified in
observations. The mechanism that we describe to burn all helium stably
occurs to a lesser extent in the regime of pure helium flashes, where
part of the helium is burned before the thermonuclear runaway \citep{Woosley2004}.
For accretion rates approaching the new stable regime, this part may
be larger.

In time-dependent multi-zone models we find that the same steady-state
regime can take place at lower $\dot{m}$, if the base heating is
stronger. We used $Q_{\mathrm{b}}=0.75\,\mathrm{MeV\,u^{-1}}$, which
is in line with the amount of heat generated by electron captures
in the crust \citep{Gupta2007}. At lower $\dot{m}$ a higher value
of $Q_{\mathrm{b}}$ is actually expected, because of the temperature
sensitivity of neutrino cooling \citep[e.g.,][]{Cumming2006}. Furthermore,
strong base heating may be needed to explain the observed recurrence
times of superbursts \citep{Keek2011} and the ignition of superbursts
in transients \citep{Keek2008,Altamirano2012}. The temperature in
the outer envelope is largely set by $\beta$CNO burning, whereas
at the bottom of the envelope crustal heating sets an inflowing flux,
such that a higher $Q_{\mathrm{b}}$ leads to a higher temperature
at the inner boundary. The temperature profile is, therefore, steeper
for the model with a higher $Q_{\mathrm{b}}$. The boundary of the
new steady-state regime is thus strongly dependent on $Q_{\mathrm{b}}$,
and a large series of time-dependent or steady-state models are required
to investigate it.

\subsection{Superburst fuel}

Assuming all energy is provided by carbon burning, cooling models
infer from superburst lightcurves a carbon mass fraction of $15\,\%-26\,\%$
\citep{Cumming2006}. Part of the energetics could originate from
photodisintegration of heavy isotopes near the \textsl{rp}-process
end-point \citep{Schatz2003ApJ}, but the composition of the ashes
is likely dominated by iron-group nuclei \citep{Woosley2004}. Unstable
burning produces at most $5\,\%$ $^{12}\mathrm{C}$ \citep[e.g.,][]{Woosley2004},
whereas the new steady-state regime yields as much as $98\,\%$: neither
gives the amount inferred for superbursts. Superbursting sources have
relatively high values of the ratio of the fluence in-between and
during bursts (the so-called $\alpha$-parameter; \citealt{Zand2004}),
which suggests that part of the accreted hydrogen and helium burns
in a stable manner and part in bursts. To obtain $20\,\%$ $^{12}\mathrm{C}$,
the steady-state mode must be active $16\,\%$ of the time. As the
steady-state occurs in a small $\dot{m}$ range, the $16\,\%$ duty
cycle may be provided by the intrinsic variations in $\dot{m}$. Alternatively,
there could be a hybrid between the bursting and stable modes, where
a substantial part of $^{4}\mathrm{He}$ burns stably as in the steady-state
regime, but an instability is triggered on a delayed timescale. This
regime would be different from the ``delayed mixed bursts'' proposed
by \citet{2003NarayanHeyl}. Further time-dependent simulations are
needed to study the transition between the different regimes, and
attention must be given to the possibility of some carbon destruction
prior to superburst ignition \citep[e.g.,][]{Keek2012}.

\section{Conclusions}

We present models of a new steady-state nuclear-burning regime for
hydrogen and helium mixtures accreted onto neutron stars. As hydrogen
burns stably via the hot CNO cycle, the increased temperature induces
$3\alpha$ burning of helium. This mechanism fully depletes helium
in steady-state at mass accretion rates of $\sim10\,\%$ of the Eddington
limit, bordered by burst regimes at lower and higher accretion rates.
Enhanced base heating can steepen the temperature profile and sustain
the steady-state regime, extending the range of accretion rates where
stable burning is active. Pure-carbon ashes are produced, which may
power the rare energetic superbursts. In fact, more carbon is produced
than needed to explain the fuel mixture of superbursts. Steady-state
burning only needs to occur $\sim16\,\%$ of the time. The new stable
burning mode could contribute to the large values of the $\alpha$-parameter
observed for superbursting sources. 

In future papers we will perform simulations to investigate the dependence
of this burning mode on base heating, metallicity, and other parameters.
Large burst catalogues such as MINBAR \citep{Keek2010} will be employed
to constrain the observational signature.

\section*{Acknowledgements}

The authors thank F.\,X.~Timmes for public availability of the Helmholtz
EOS and opacity routines at \url{http://cococubed.asu.edu}, and they
are grateful for discussions with A.~Cumming and E.\,F.~Brown.
LK acknowledges support from NASA ADAP grant NNX13AI47G. AH is supported
by an ARC Future Fellowship (FT120100363). This material is based
upon work supported by the National Science Foundation under Grant
No. PHY-1430152 (JINA Center for the Evolution of the Elements).

\bibliographystyle{mnras}
\bibliography{interstable1}

\bsp	
\label{lastpage}
\end{document}